# Visible Insights of the Invisible Pandemic: A Scientometric, Altmetric and Topic Trend Analysis

**Sujit Bhattacharya[1] and Shubham Singh**

CSIR-National Institute of Science Technology and Development Studies
Pusa Campus, K.S. Krishnan Marg, New Delhi, India

## Abstract

The recent SARS-COV-2 virus outbreak has created an unprecedented global health crisis! The disease is showing alarming trends with the number of people getting infected with this disease, new cases and death rate are all highlighting the need to control this disease at the earliest. The strategy now for the governments around the globe is how to limit the spread of the virus until the research community develops treatment/drug or vaccination against the virus. The outbreak of this disease has unsurprisingly led to huge volume of research within a short period of time surrounding this disease. It has also led to aggressive social media activity on twitter, Facebook, dedicated blogs, news reports and other online sites actively involved in discussing about the various aspects of and related to this disease.

It becomes a useful and challenging exercise to draw from this huge volume of research, the key papers that form the research front, its influence in the research community, and other important research insights. Similarly, it becomes important to discern the key issues that influence the society concerning this disease. The paper is motivated by this. It attempts to distinguish which are the most influential papers, the key knowledge base and major topics surrounding the research covered by COVID-19. Further it attempts to capture the society's perception by discerning key topics that are trending online. The study concludes by highlighting the implications of this study.

**Keywords:** COVID-19, Pandemic, Social Distancing, Altmetrics, Dimensions database, Co-citation analysis. Google Trends

## Introduction

Coronaviruses are viruses that circulate among animals and are named because of the crown-like spikes (protein spikes) that protrude from their surface resembling the sun's corona. The first transmission of this type of virus from animals to humans happened in 2002 in the Guangdong province of China which resulted in SARS (Severe Acute Respiratory Syndrome). Bats were thought to be the potential source of this virus. A novel coronavirus (nCoV) was identified in early January 2020 which was traced to the severe pneumonic outbreak of an undocumented cause in early December 2019 in the city of Wuhan, China. Similar to SARS, bats are seen as the potential source from which this virus has spread to humans. Due to rapid spread of the virus within a short time globally resulting in health emergencies in a number of

[1] For correspondence: sujit@academic@yahoo.com

countries, WHO declared this as a pandemic on 11th March 2020. Initially the virus was named 2019-nCoV but later was named SARS-CoV-2 due to its "genetic relationship to the SARS-COV-1 virus"(Medscape, 2020)

The entire human population is at potential risk as being a new virus nobody has prior immunity to it. There is no vaccine and no specific treatment for the disease and is highly transmissible. Epidemiological estimation at present is that on an average, one infected person will infect between two to three other people. The spread has been from respiratory droplets and from contaminated surfaces. The world is facing a common challenge; how to control the spread of COVID-19 and what can be the effective interventions to control mortality. The early examples from China suggested the most effective method to control the spread of the virus is through lockdowns and social distancing measures. The example of South Korea suggested testing as the major component of the mitigation measures. Some studies pointed out the importance of hand hygiene and face masks in controlling the virus. With new hotspots emerging, the number of new cases and those not able to recover are raising new concerns every day. Risk and uncertainty behind this disease control has generated a global concern for health, economy, and for persons at large. The alarming spread of the virus has shocked people across the world pushing among others researchers to understand the virus—its structure, transmission, replication mechanism, latency, etc. and promising interventions that can effectively control it. Extensive global efforts are undertaken to develop vaccine and drug. This is unsurprisingly leading to huge volume of research activity within a short period of time increasing at an exponential rate. As the recent editorial published in *The Lancet* highlights *"The whole-genome sequence of SARS-CoV-2 had been obtained and shared widely by mid-January, a feat not possible at such speed in previous infectious disease outbreaks"* The editorial points out the importance of the need for development of effective diagnostics, therapeutics and vaccine for the virus. *Examining from the Dimensions database as of 22 April, it was found that 1633 clinical trials are being conducted on the virus and 171 policy documents have been published so far.* The number of research papers, clinical trials at different phases within such a short period is unprecedented and shows the intensive efforts of the global research community to understand the different aspects of this disease and address it. *Seven patents have also been granted.* It is important to capture insights of influential research and innovation from this ongoing activity for policy makers and research scholars from cross-disciplinary areas to build up further on this valuable repository. Societal impact and what aspects are of concern to the people at large are difficult to capture. One useful method would be from online trends surrounding this disease that would indicate to some extent the key issues that are influencing the society at large. The present study is motivated by this and applies tools and techniques of scientometrics to uncover insights from research papers.

Scientometrics applies various mathematical and statistical techniques to capture insights of research activity from research papers and patents and other published sources including online sources (Altmetrics) by constructing various types of indicators. Citation based analysis is a prominent method to capture academically significant and theoretically relevant material (see for example Glanzel, 2003). *Keeping in view the research activity in this area started primarily with the outbreak of this disease, impact captured through citations would not give a correct picture as citations takes time to accrue.* This is true for research paper as well as patent citations. Citations that influence current research activity would however be useful to construct the present knowledge base. One of the useful method that can do so is based on co-citation analysis which captures frequency with which two documents are cited together (Small, 1973). Co-Citation establishes an intellectual relationship with earlier literature in a field/subfield/area of research; strength of relationship based on frequency of co-citation pairs. The rise of social networking websites like twitter, Facebook etc. provides researchers a wider scope to share their scholarly publications. Altmetrics allows to track and capture online impact of scholarly research and thus broadly indicates papers that are influencing the research community. To put it in a proper perspective, one can borrow from William (2017), "Altmetrics are measurements of how people interact with a given scholarly work". Altmetrics or article level metrics according to Das and Mishra (2014) is a "new trendsetter" to measure "impact of scientific publication and their social outreach to intended audience". It reflects "a scholarly article's popularity, usage, acceptance and availability" by using an altmetric score.

Google Trends which was launched in 2006, primarily shows how frequently a particular search term is entered in comparison with all other search terms in different regions and languages (Google, 2017). In Google trends level of interest in a topic is approximated using search volume of Google. Sullivan (2016) estimated searches on Google Trends reached 2 trillion in 2016! Thus, this is one of the most significant source of data if it is properly analysed. One of the most influential study was by Ginsberg et al. (2006) which showed that Google Trends traced and predicted the spread of influenza earlier than the Centers for Disease Control and Prevention. Jun *et al.* (2018) provides a good assessment of research studies in the past decade which have utilized Google Trends. They highlight the diverse fields in which this has been used for, from merely describing and diagnosing research trends to forecasting changes. According to Mavragani et al. (2018) "Google Trends shows the changes in online interest for time series in any selected term in any country or region over a selected time period, for example, a specific year, several years, 3 weeks, 4 months, 30 days, 7 days, 4 hours, 1 hour, or a specified time-frame." They argue that as the internet penetration is increasing web based search activity has become a valid indicator of public behaviour.

The paper positions itself in this direction; applying various tools and techniques of scientometrics, Altmetrics and Google Trends to draw meaning from the huge volume of research papers and online activity surrounding this pandemic. The study attempts to answer the following research questions:

- What are the key papers that captures the most relevant research, areas and topics on COVID-19?
- What is the knowledge base that influences current research on this pandemic?
- What are the key aspects of this pandemic that is influencing the society at large?

**Method:** The study has used various types of data sets and analytical techniques as highlighted below to capture the research trends and also assess this disease influence on the society. The Dimensions database (www.dimensions.ai) was used for this study. This database has various unique features which makes it very useful to capture various aspects of research activity. It provides dynamic Altmetrics score for each article. The database unlike source based classification (journal classification) used in indexing articles in SCI and Scopus database uses article level classification. Only when an article cannot be classified individually due to lack of information, it uses the Fields of Research (FOR) classification system. The FOR[2] has three hierarchical levels: Divisions (represents a broad subject area or research discipline), with the next two levels Groups and Fields representing increasingly detailed subsets of these categories. In FOR there are 22 Divisions, 157 Groups and 1238 Fields. Dimensions has incorporated only the Groups in its classification system. Thus classification article level provides a more informed assessment of the topic covered by it then based on journal level classification which is a macro level classification.  These features motivated us to use this database for this study.

The articles on this virus were extracted using the search string "Covid-19" Or "SARS-CoV-2" Or “SARS-CoV2” Or "2019-nCoV" on April 12, 2020 from this database. The final search string was developed based on review of contemporary studies and deleting those search keywords that lead to noises. For example, it was found that nCoV which some studies have used also identifies papers that cover MERS (Middle-East respiratory syndrome). This was first reported in 2012, was initially called novel coronavirus or nCoV as it was a species of coronavirus. Many studies had applied search string without hyphen which also results in extracting papers not covering this disease. The search string applied on the publications database of Dimensions resulted in 9146 papers, containing 7332 articles and 1814 pre-prints. This data set of 9146 papers were further used for analysis. Influential papers were distinguished by using Altmetric score which is a weighted count of all the online attention of a research paper. The altmetrics data was captured from Dimensions database which draws data from altmetrics.com of capturing online activity of research papers on Facebook, twitter, blogpost, news reports etc. The score changes as people mentioning the paper increases (only one mention per user is considered). Each category of mention carries different base amount so a news article contributes more than a blog post which in turn contributes more than a tweet in the final score.

Country wise analysis showed that around 78 percent of the total papers were contributed by ten countries. Further analysis of research activity of the ten identified countries was done using altmetric and citation analysis. Word cloud provides a high visual representation of concepts that a paper had frequently applied. It is based on Burst algorithm that captures the sudden rise in the usage of a word. Mane and Borner (2004) highlighted the usefulness of burst words as according to them “it helps humans mentally organize and electronically access and manage large complex information spaces”. Using R programming tools, word cloud was constructed from keywords of the data set. The words with higher frequency in the overall corpus of papers (herein 9146) have a larger font size and acquires more space in a visualisation. Word Cloud was used to get visualisation of 70 most frequent words; the number of words chosen was limited by the clarity of visualisation.

Co-citation analysis helps to capture papers that are co-cited together in a large number of papers. The highly co-cited papers is seen as the core knowledge base of research area at a particular period.  This analysis was undertaken to identify the key knowledge base behind the identified papers. Dimensions database was used to extract a bibliographic mapping file for the papers. Two software's Pajek and VOSviewer were used for co-citation analysis. Initially the bibliographic mapping file was run on the VOSviewer software to identify the most co-cited papers. The co-cited papers were identified at four levels (trim levels) to have a deeper insight of the core knowledge base: Level 1 identified 51 papers co-cited 77 or more times; Level 2 identified 26 papers co-cited 126 or more times; Level 3 identified 10 papers co-cited 277 or more times, and Level 4 identified 5 papers co-cited 463 or more times. For each of the trim levels a network file was obtained from the VOSviewer software. The network file was then run on Pajek software to create a refined co-citation network map so as to avoid overlapping of nodes. The final visualisation was done for the refined map in the VOSviewer software. Data for the Policy Documents referencing these top ten co-cited papers was done by accessing altmetrics.com directly from the dimensions database.

---

[2]Is a  component of the Australian and New Zealand Standard Research Classification (ANZSRC) https://www.abs.gov.au/ausstats/abs@.nsf/Latestproducts/1297.0Contents12008?opendocument&tabname=Summary&prodno=1297.0&issue=2008&num=&view=

Another question which the study explored is the impact of this virus on the society primarily what are the key aspects of and related to this disease that has influenced the society at large. Google trend analysis of key topics have been undertaken to capture this aspect. Google trends website (https://trends.google.com/trends/?geo=US) was first accessed on 10th April 2020. The topics were chosen based on closely monitoring the news items, and also finally choosing from a large set of topics. Choice for example of 'pandemic' was seen to have initial burst but declined quickly. Vaccine was trending highly but we found lot of noise in this term. The final six topics chosen were "Social distancing", "Quarantine", "COVID-19", "Coronavirus", "Face Mask" and "Hydroxychloroquine". Data for each of the topics was finally taken on 16 April, 2020. For country specific comparison data for five countries having maximum cases of COVID-19 namely USA, Italy, Spain, France, Germany and two emerging economies India and Brazil was also obtained. Hydroxychloroquine was not used in country specific search as it was only visible trending for three countries among the chosen seven countries. Google trend analysis was not done for China as there is much restricted access to Google in that country.

## Results

### Insights from COVID-19 research papers

One of the first important observation is the intensity with which research on COVID-19 and related aspects is going on globally. Search conducted in two different time periods, 28th March and then on 12th April showed that 2172 and 9146 papers were published in these two periods; almost 320 percent growth during such a short time.

The insights that we draw from our analysis of the 9146 papers is presented in different sections below

### Popular research papers: Altmetrics and Content analysis

Ten most popular research papers among the 9146 COVID-19 papers were extracted on the basis of their Altmertrics score on April 12, 2020. Table 1 highlights these influential papers.

**Table 1: Ten most popular research papers on COVID-19**

| Paper | Research Area | Altmetrics Score | Popularity Breakup |
|---|---|---|---|
| Andersen, G.K, Rambaut, A., Lipkin, W.I., Holmes, E.C., & Garry, R.F. (2020). Proximal Origins of SARS-COV-2. *Nature Medicine*, https://doi.org/10.1038/s41591-020-0820-9 | Medical and Health Sciences | 27718 | • Twitter- 74942 tweets from 67349 users<br>• News- 541 news stories from 353 outlets<br>• Blog- 58 Posts from 47 blogs<br>• Facebook- 49 public wall posts from 47 users |
| van Doremalen, N., Bushmaker, T., Morris, D.H., Holbrook, M.G., Amandine Gamble, A., et al. (2020). Aerosol and Surface Stability of SARS-CoV-2 as Compared with SARS-CoV-1. *New England Journal of Medicine*, doi: 10.1056/NEJMc2004973 | Medical and Health Sciences | 21877 | • Twitter- 23842 tweets from 20633 users<br>• News- 1573 news stories from 722 outlets<br>• Blog- 111 posts from 77 blogs<br>• Facebook- 33 public wall posts from 29 users<br>• 6 Policy Documents reference this paper |
| Li, R., Pei, S., Chen, B., Song, Y., Zhang, T., Yang, W., & Shaman, J. (2020). Substantial undocumented infection facilitates the rapid dissemination of novel coronavirus (SARS-CoV2). *Science*, doi: 10.1126/science.abb3221 | • Medical and Health Sciences<br>• Medical Microbiology | 15634 | • Twitter- 22965 tweets from 19798 users<br>• News- 485 news stories from 293 outlets<br>• Blog- 61 posts from 49 blogs<br>• Facebook- 15 public wall posts from 14 users<br>• 2 Policy documents reference this paper |
| Leung, N. H. L., Chu, D. K. W., Shiu, E. Y. C., Chan, K.-H., Mcdevitt, J. J., Hau, B. J. P. et al. (2020). Respiratory virus shedding in exhaled breath and efficacy of face masks. *Nature Medicine*. doi: 10.1038/s41591-020-0843-2 | • Medical and Health Sciences<br>• Medical Microbiology | 15370 | • Twitter- 34608 tweets from 29664 users<br>• News- 207 news stories from 149 outlets<br>• Blog- 35 posts from 32 blogs<br>• Facebook- 5 public wall posts from 5 users |
| Shen, C., Wang, Z., Zhao, F., Yang, Y., Li, J., Yuan, J. (2020). Treatment of 5 Critically Ill Patients With COVID-19 With Convalescent Plasma. Jama. doi: 10.1001/jama.2020.4783 | • Medical and Health Sciences<br>• Clinical Sciences | 14379 | • Twitter- 43514 tweets from 39886 users<br>• News- 179 news stories from 150 outlets<br>• Blog- 17 posts from 15 blogs<br>• Facebook- 5 public wall posts from 5 users |
| Caly, L., Druce, J. D., Catton, M. G., Jans, D. A., & Wagstaff, K. M. (2020). The FDA-approved Drug Ivermectin inhibits the replication of SARS-CoV-2 in vitro. *Antiviral Research*, 104787. doi: 10.1016/j.antiviral.2020.104787 | • Medical and Health Sciences<br>• Medical Microbiology | 11630 | • Twitter- 17977 tweets from 15077 users<br>• News- 257 news stories from 221 outlets<br>• Blog- 12 posts from 12 blogs<br>• Facebook- 7 public wall posts from 7 users |
| Wu, Z., & McGoogan, J.M. (2020). Characteristics of and Important Lessons From the Coronavirus Disease 2019 (COVID-19) Outbreak in China. *JAMA*, doi:10.1001/jama.2020.2648 | Medical and Health Sciences | 11283 | • Twitter- 17896 tweets from 15509 users<br>• News- 991 news stories from 426 outlets<br>• Blog- 88 posts from 54 blogs<br>• Facebook- 49 public wall posts from 45 users |

| Paper | Research Area | Altmetrics Score | Popularity Breakup |
|---|---|---|---|
| | | | • 6 Policy Documents reference this paper |
| Lauer, S.A., Grantz, K.H., Bi, Q., Jones, F.K., Zheng, Q., Meredith, H.R., et al. (2020). The Incubation Period of Coronavirus Disease 2019 (COVID-19) From Publicly Reported Confirmed Cases: Estimation and Application. *Annals of Internal Medicine*, doi: https://doi.org/10.7326/M20-0504 | • Medical and Health Sciences<br>• Public Health and Health Services | 8090 | • Twitter- 4823 tweets from 4484 users<br>• News- 1026 news stories from 561 outlets<br>• Blog- 47 posts from 39 blogs<br>• Facebook- 11 public wall posts from 10 users<br>• 1 Policy Document referenced this paper |
| Cao, B., Wang, Y., Wen, D., Liu, W., Wang, J., Fan, G., et al. (2020). A Trial of Lopinavir–Ritonavir in Adults Hospitalized with Severe Covid-19. *New England Journal of Medicine*, doi: 10.1056/NEJMoa2001282 | • Medical and Health Sciences<br>• Clinical Sciences | 7746 | • **Twitter- 9036 tweets from 7931 users**<br>• **News- 299 news stories from 198 outlets**<br>• **Blog- 40 posts from 37 blogs**<br>• **Facebook- 24 public wall posts from 23 users**<br>• **1 Policy Document referenced this paper** |
| Fang, L., Karakiulakis, G., & Roth, M. (2020). Are patients with hypertension and diabetes mellitus at increased risk for COVID-19 infection? *The Lancet Respiratory Medicine*, DOI:https://doi.org/10.1016/S2213-2600(20)30116-8 | • Medical and Health Sciences<br>• Clinical Sciences<br>• Public Health and Health Services | 6902 | • **Twitter- 7108 tweets from 6154 users**<br>• **News- 307 news stories from 220 outlets**<br>• **Blog- 30 posts from 28 blogs**<br>• **Facebook- 18 public wall posts from 18 users** |

Note: *Paper Pradhan, "P., Pandey, A. K., Mishra, A., Gupta, P., Tripathi, P. K., Menon, M. B. et al. (2020). Uncanny similarity of unique inserts in the 2019-nCoV spike protein to HIV-1 gp120 and Gag. doi: 10.1101/2020.01.30.927871" with 14263 altmetric score was also one of the popular paper. This paper is not given here because it has been withdrawn by the authors subsequently*

Content analysis of the ten papers gives us a broad idea of the papers and some insights of why they have become popular. Anderson *et al. (2020)*, the study most popular on social media platforms (number of tweets more than three times the next popular paper) commented that "SARS-COV-2" is the seventh coronavirus to infect humans". The study found that SARS-COV-2 is not a product of purposeful manipulation and is most likely the result of natural selection of human or human-like ACE2 receptor. The study also found that SARS-COV-2 spike protein has high affinity to bind to human ACE2 receptor. *Issues of how the human body binds to the virus is an active area of research which thus has attracted so much attention*. van Doremelen *et al. (2020)* analysed the "aerosol and surface stability of SARS-COV-2 and compared it with SARS-COV-1. The study highlighted the need to protect from five environmental conditions: aerosols, plastic, stainless steel, copper and cardboard. Further it showed that SARS-COV-2 remains viable in aerosols for 3 hours, on plastic and stainless steel for 3 days, 4 hours on copper and on cardboard for 24 hours. The study found that the stability of SARS-COV-1 is similar on plastic, stainless steel and aerosols to SARS-COV-2 and different to SARS-COV-2 on cardboard (8 hours) and copper (8 hours). *The implications of this study can be clearly seen in the preventive measures of COVID-19. This study was also cited in policy documents.* Li *et al. (2020)* estimated that 86% of COVID-19 cases went undocumented in China prior to their travel restrictions. The study also estimated that the undocumented cases contagiousness or transmission rate was 55% of documented infections, yet 79% of documented infection cases were due to these undocumented infections. The suggestion of this study that undocumented infections "isolation and identification is necessary to fully control the virus" is very important and the spread of this virus may be seen as a consequence of this. This study also was cited in policy documents.

Leung *et al.* (2020) explored "the importance of respiratory droplet and aerosol route of transmission" by quantifying the "amount of respiratory virus in exhaled breath of participants" that have acute respiratory virus illness (ARI). The 246 participants were divided in two groups, one wearing surgical face mask and other not wearing face mask. The study found that surgical face masks can efficaciously reduce the respiratory droplet emission of influenza virus particles but not in aerosols. They also found that surgical face masks can be used by ill patients of COVID-19 to reduce "onward transmission". Face mask is getting increasing attention and now being incorporated as essential guideline in health policies of different countries. *The paper provided a good empirical support to this i.e. face masks*. Shen *et al.* (2020) study examined covalescent plasma transfusion benefit in treatment of critically ill COVID-19 patients. The clinical trial was conducted on 5 critically ill patients with COVID-19 and acute respiratory distress syndrome (ARDS) along with certain other conditions. The study found decline in viral load, clinical conditions of patients improved as "indicated by body temperature reduction, improved $PAO_2/FIO_2$ and chest imaging". This treatment is now being incorporated in many countries. *This study is first of its kind in this direction and is thus not surprising that it is one of the most popular research paper in COVID-19*.

Caly *et al.* (2020) tested Ivermectin's (an FDA approved drug) antiviral activity towards SARS-COV-2. The study found that a single dose of Ivermectin was "able to control viral replication within 24-48 hours" in the system provided by them. Doctors are struggling to control this dangerous disease. *A study like this which gives some hope is quickly tracked which is indicated by its high altmetrics score*. Wu and McGoogan (2020) summarised the key findings of Chinese Center for Disease Control and Prevention. The study found that 81% of cases were mild with no deaths, 14% were severe with no deaths and 5% were critical cases with 49% Case Fatality Rate (CFR). The study also found that patients with comorbid conditions like cardiovascular diseases and diabetes had 10.5% and 7.3% CFR respectively. The disease started in Wuhan, China and within a short time a large number of people were affected by it. *The popularity of this study can be seen from the insights emerging from this study that provides a good empirical estimation of the effects of this disease. The fact that policy documents are also referencing this paper also substantiates this claim.*

Lauer *et al.* (2020) estimated "the length of incubation period of COVID-19 and described its public health implications". Data of 181 cases that tested positive for SARS-CoV-2 infection outside Hubei province was taken for the study. The study estimated 5.1 days to be the median incubation period of COVID-19. It also found that patients will develop symptoms within 14 days. There is however chance of 1-2 percent of patients developing symptoms after 14 days. The upper bound of 14 days thus has been proposed by the study to be the length of quarantine and active monitoring. *The study thus provided a good estimation of the period of quarantine and thus it is getting high attraction from scholars is not surprising.* The paper is also cited by WHO in a policy document.

Cao *et al. (2020)* analysed 199 patients with positive SARS-CoV-2. The patients were assigned in a 1:1 ratio and 99 patients received lopinavir-ritonavir twice daily with standard care and 100 patients received only standard care (comprised of ventilation, supplemental oxygen etc) for 14 days. The study found that lopinavir-ritonavir was not able to “accelerate clinical improvement, reduce mortality or diminish throat viral RNA detectability in patients with serious COVID-19”. *The study is important as it cautions doctors towards administrating this drug for COVID-19.* The study by Fang *et al. (2020)* highlighted that ACE2 (angiotensin-converting enzyme 2) expression is increased in patients with diabetes, and hypertension. The study underscored the reasons behind it to the treatment being administered to them; these patients are treated with ACE inhibitors and ARBs (angiotensin II type-I receptor blockers) which increases the expression of ACE2. As the SARS-CoV-2 binds to target cells via ACE-2 the study thus concludes that infection is facilitated in patients with these conditions. *The importance of the study can be seen in the case of making aware, protecting and preventing the spread of COVID-19 to these vulnerable populations. This has largely contributed to its popularity.*

**Table 2: Ten Most Active Countries in COVID-19 Research**

| Country | Papers | Key Organisations (Papers) | Paper with Highest Citation (Citations) | Most Popular Paper (Altmetric Score) | Key Collaborations (Papers) |
|---|---|---|---|---|---|
| China | 2,543 | • Huazhong University of Science and Technology (HUST) (152)<br>• University of Hong Kong (120) | Huang et al. (2020), *The Lancet* (1200) | Li et al. (2020), *Science* (15467) | USA (363)<br>UK (131) |
| USA | 1,740 | • Harvard University (94)<br>• Johns Hopkins University (JHU) (64) | Holshue et al. (2020), *NEJM* (279) | Anderson et al (2020), *Nature Medicine* (27667) | China (363)<br>UK (158) |
| UK | 781 | • University of Oxford (72)<br>• LSHTM (62) | Wang et al. (2020), *The Lancet* (173) | Anderson et al (2020), *Nature Medicine* (27667) | USA (158)<br>China (131) |
| Italy | 501 | • University of Milan (UNIMI) (55)<br>• Sapienza University of Rome (45) | Hui et al. (2020), *Int. Journal of Infectious Disease* (158) | Remuzzi and Remuzzi, (2020), *The Lancet* (5846) | US (104)<br>UK (69) |
| Germany | 313 | • Charité – University Medicine Berlin (19)<br>• Heidelberg University (13) | Rothe et al. (2020), *NEJM* (254) | Kampf et al. (2020), *Journal of Hospital Infection* (12192) | US (131)<br>China (60) |
| Canada | 280 | • University of Toronto (76)<br>• University of British Columbia (UBC) (40) | Jin et al. (2020), *Military Medical Research* (55) | Emanuel et al. (2020), *NEJM* (4719) | US (85)<br>China (59) |
| France | 264 | • Méditerranée Infection Foundation (18)<br>• University of Bordeaux (10) | Corman et al. (2020), *Eurosurveillance* (124) | Gautret et al. (2020), *Int. Journal of Antimicrobial Agents* (6563) | China (68)<br>US (62) |
| Australia | 252 | • University of Sydney (USYD) (37)<br>• Monash University (32) | Lu et al (2020), *The Lancet* (373) | Anderson et al (2020), *Nature Medicine* (27667) | China (84)<br>US (73) |
| India | 220 | • Dr. D.Y. Patil Vidyapeeth, Pune (22)<br>• AIIMS (15) | Bannerjee (2020), *Asian Journal of Psychiatry* (18) | Pradhan et al. (2020), bioRxiv (14261) | China (38)<br>US (31) |
| Switzerland | 174 | • University of Zurich (UZH) (18)<br>• University Hospital of Basel (USB) (13) | Rious and Althaus (2020), *Eurosurveillance* (77) | Fang et al. (2020), *The Lancet Respiratory Medicine* (6913) | US (37)<br>Germany (33) |

Table 2 points to some interesting aspects of research activity in this area. These ten countries account for almost 78 percent of total papers with China and USA accounting for 45 percent of the total. China, USA and UK are actively collaborating among themselves and also with other countries. This is a good indication as global collaborative efforts, pooling each other resources are required to meet the challenges posed by this disease. A few leading universities can be discerned which are actively involved in this research. Popularity of a paper can also be seen influenced by journals; papers with high altmetrics score strongly correlate with journals that have high reputation in the field (high impact factor, leading journal of the community).

Table 3 highlights the areas covered in the COVID-19 papers.

**Table 3: Key areas in Covid-19 papers***

| **Name** | **Fields of Research code** | **Publications** | **Citations** |
|---|---|---|---|
| Medical and Health Sciences | 11 | 5741 | 19145 |
| Public Health and Health Services | 1117 | 2355 | 5782 |
| Clinical Sciences | 1103 | 2038 | 6713 |
| Medical Microbiology | 1108 | 908 | 4854 |
| Biological Sciences | 06 | 577 | 2141 |
| Cardiorespiratory Medicine and Haematology | 1102 | 281 | 561 |
| Biochemistry and Cell Biology | 0601 | 261 | 704 |
| Immunology | 1107 | 213 | 217 |
| Information and Computing Sciences | 08 | 184 | 35 |

***Total papers 9146 papers; Source:** Dimension Database

The table provides a broad indication of intensity of research happening in different fields.

Figure 1 presents a word cloud of most frequently used terms in Covid-19 Papers.

**Figure 1: Word Cloud of Documents of “Covid19” Papers**

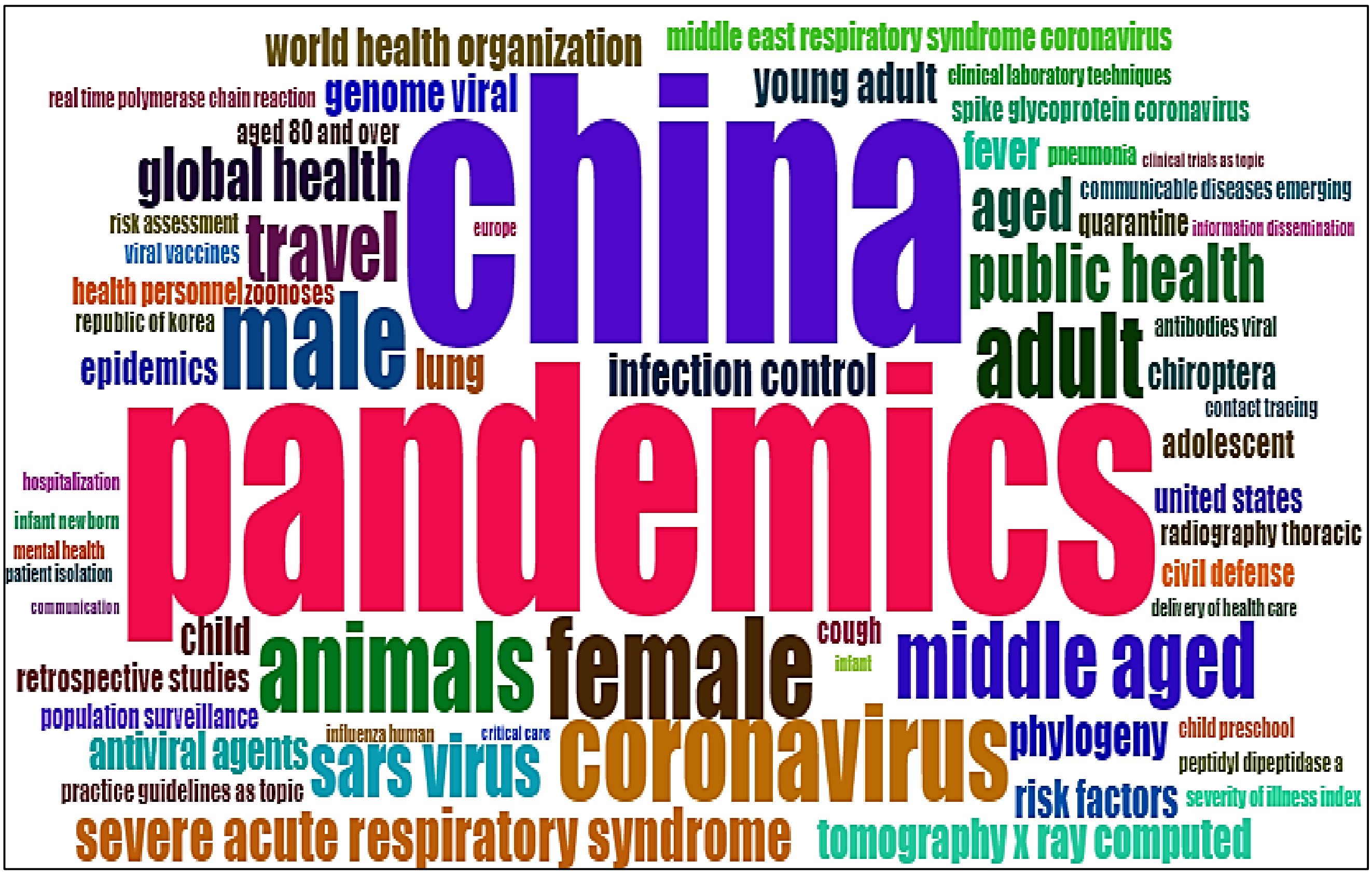

The word cloud shows key aspects that have been part of many studies. *The word cloud maps the topics of research surrounding this disease.* The two keywords, for example “Pandemics” and “China” that have maximum occurrence in papers indicated by large font size which shows that these two aspects were discussed in many papers. It is known that China was the source of this infection and WHO declared this disease as a pandemics. Thus increasing research mention of these two keywords is not surprising. Coronavirus primarily affects animals, SARS disease as a result of transmission of coronavirus from animal to humans, travel has contributed maximum to the spread of this disease, are all visible prominently in this word cloud. Thus, examination of the word cloud is useful to have a broad view of key areas of research in the 9146 papers.

**Co-Citation Analysis of COVID-19 Papers**

Figure 2 shows co-citation networks at four trim levels.

**Figure 2: Co-Citation Networks at various trim levels**

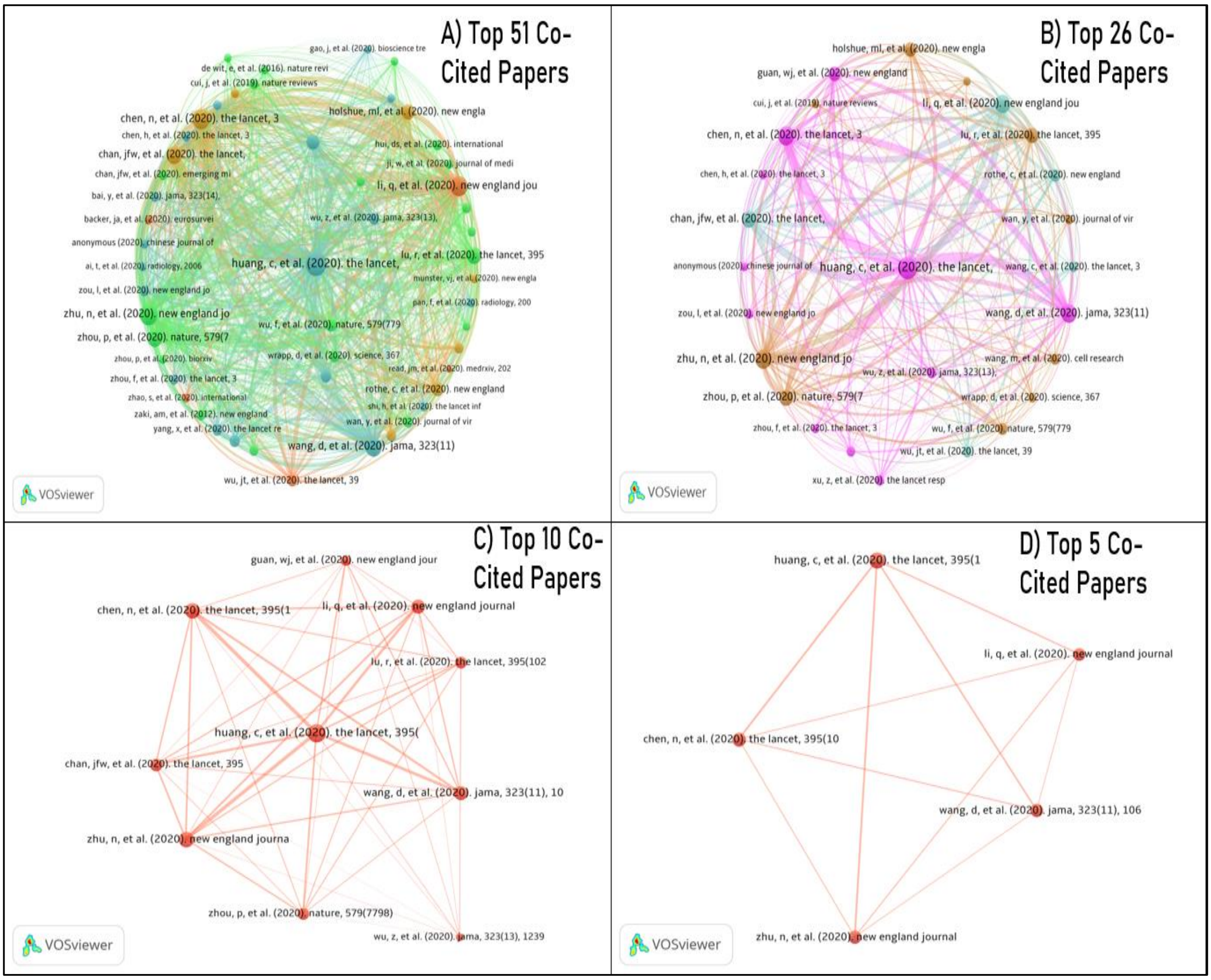

It can be observed from figure 2 that Trim level 4 that contains top 5 co-cited papers is a complete cluster. A complete cluster according to GMÜR (2003) is when *"each reference is connected to other references and there is no dominant document within the cluster"*.

Table 4 highlights the details of these co-cited papers at trim level 3 which identifies top 10 co-cited papers. It also includes the top 5 co-cited paper at Trim level 4 (refer methodology for details).

## Table 4: Top 10 Co-Cited Papers

| Paper* | Co-Citations (Total Link Strength) | Total Citations | Key Areas of Citing Papers (%share) | Organisation (Policy Documents Citing this Paper) | Journal Impact Factor |
|---|---|---|---|---|---|
| Huang, C., Wang, Y., Li, X., Ren, L., Zhao, J., Hu, et al. (2020). Clinical features of patients infected with 2019 novel coronavirus in Wuhan, China. *The Lancet*, *395*(10223), 497–506. doi: 10.1016/s0140-6736(20)30183-5 | 1150 (21356) | 1200 | • Medical and Health Sciences (86.89)<br>• Biological Sciences (7.71) | • WHO (15)<br>• CDC (2)<br>• Others (2) | 59.102 |
| Zhu, N., Zhang, D., Wang, W., Li, X., Yang, B., Song, J., et al. (2020). A Novel Coronavirus from Patients with Pneumonia in China, 2019. *New England Journal of Medicine*, 382(8), 727–733. doi: 10.1056/nejmoa2001017 | 766 (15179) | 787 | • Medical and Health Sciences (83.03)<br>• Biological Sciences (11.27) | • CDC (2) | 70.670 |
| Chen, N., Zhou, M., Dong, X., Qu, J., Gong, F., Han, Y., et al. (2020). Epidemiological and clinical characteristics of 99 cases of 2019 novel coronavirus pneumonia in Wuhan, China: a descriptive study. *The Lancet*, 395(10223), 507–513. doi: 10.1016/s0140-6736(20)30211-7 | 620 (12311) | 647 | • Medical and Health Sciences (89.58)<br>• Biological Sciences (5.37) | • WHO (10)<br>• UK Govt. (1)<br>• Scottish Govt. (1) | 59.102 |
| Zhou, P., Yang, X.-L., Wang, X.-G., Hu, B., Zhang, L., Zhang, W., et al. (2020). A pneumonia outbreak associated with a new coronavirus of probable bat origin. *Nature*, 579(7798), 270–273. doi: 10.1038/s41586-020-2012-7 | 453 (11820) | 461 | • Medical and Health Sciences (72.46)<br>• Biological Sciences (20.70)<br>• Technology(1.66) | | 43.070 |
| Li, Q., Guan, X., Wu, P., Wang, X., Zhou, L., Tong, Y., et al. (2020). Early Transmission Dynamics in Wuhan, China, of Novel Coronavirus–Infected Pneumonia. *New England Journal of Medicine*, 382(13), 1199–1207. doi: 10.1056/nejmoa2001316 | 685 (11485) | 698 | • Medical and Health Sciences (86.25)<br>• Biological Sciences (6.22)<br>• Mathematical Sciences (1.88) | • WHO (10)<br>• UK Govt. (1)<br>• Scottish Govt. (1) | 70.670 |
| Wang, D., Hu, B., Hu, C., Zhu, F., Liu, X., Zhang, J., et al. (2020). Clinical Characteristics of 138 Hospitalized Patients With 2019 Novel Coronavirus–Infected Pneumonia in Wuhan, China. *Jama*, 323(11), 1061. doi: 10.1001/jama.2020.1585 | 614 (11340) | 627 | • Medical and Health Sciences (93.49)<br>• Biological Sciences (4.40) | • WHO (5)<br>• UK Govt. (1) | 51.273 |
| Lu, R., Zhao, X., Li, J., Niu, P., Yang, B., Wu, H., et al. (2020). Genomic characterisation and epidemiology of 2019 novel coronavirus: implications for virus origins and receptor binding. The Lancet, 395(10224), 565–574. doi: 10.1016/s0140-6736(20)30251-8 | 366 (9722) | 373 | • Medical and Health Sciences (76.18)<br>• Biological Sciences(16.34)<br>• Technology (2.49) | | 59.102 |
| Chan, J. F.-W., Yuan, S., Kok, K.-H., To, K. K.-W., Chu, H., Yang, J., et al. (2020). A familial cluster of pneumonia associated with the 2019 novel coronavirus indicating person-to-person transmission: a study of a family cluster. *The Lancet*, 395(10223), 514–523. doi: 10.1016/s0140-6736(20)30154-9 | 437 (9058) | 455 | • Medical and Health Sciences (83.52)<br>• Biological Sciences (10.53)<br>• Psychology and Cognitive Sciences (1.14) | • WHO (10)<br>• CDC (3)<br>• Scottish Govt. (1) | 59.102 |
| Guan, W.-J., Ni, Z.-Y., Hu, Y., Liang, W.-H., Ou, C.-Q., He, J.-X. (2020). Clinical Characteristics of Coronavirus Disease 2019 in China. New England Journal of Medicine. doi: 10.1056/nejmoa2002032 | 423 (7445) | 431 | • Medical and Health Sciences (92.12)<br>• Biological Sciences (3.10)<br>• Mathematical Sciences (1.43) | • CDC (9)<br>• WHO (3)<br>• Others (2) | 70.670 |
| Holshue, M. L., Debolt, C., Lindquist, S., Lofy, K. H., Wiesman, J., Bruce, H., et al. (2020). First Case of 2019 Novel Coronavirus in the United States. New England Journal of Medicine, 382(10), 929–936. doi: 10.1056/nejmoa2001191 | 270 (6760) | 279 | • Medical and Health Sciences (87.55)<br>• Biological Sciences (7.39)<br>• Technology (1.17) | • CDC (4)<br>• WHO (2) | 70.670 |

**The articles are ranked based on total link strength* 

Table 4 in a sense identifies the proximity of papers in “terms of content” (GMÜR, 2003). According to Trujillo and Long (2018) the results of document co-citation analysis shows the “peer-recognised” literature via their “citation patterns”. This analysis has provided us the scope of mapping the COVID-19 literature. It can be observed that studies relating to epidemiology and clinical characteristics of COVID-19 have the greatest Co-citation frequency. Strength of links of Huang *et al.* (2020) paper was the highest and is the most referenced paper in key Policy documents.

**Pandemic’s Influence on the Society**

Figure 3 provides global Google trends of six topics (refer methodology for details) currently talked about extensively on social media, news reports or in general public discussion. It can be observed the otherwise flat line of COVID-19 started seeing spikes in late February, 2020. This is because the term came into existence when WHO on Feb 11 named the disease from the virus as COVID-19 (Coronavirus Disease 2019). It can also be observed that the term reached maximum level of interest during the end of March as cases started showing significant increase in countries USA, India etc. Measures like quarantine has strong societal influence and thus useful to look at trend in this topic. Another topic “Pandemic” saw maximum interest around March 11, as WHO announced COVID-19 pandemic on that day. It can also be seen that the interest about this topic fell shortly thereafter. As discussed in the methodology, this topic hence was not chosen further in this study for Google trend examination. Hydroxychloroquine term has seen a great amount of interest from middle of March, 2020. This can be traced to the study by reputed French physician and microbiologist Didier Raoult who highlighted the use of this antimalarial drug in the treatment of this disease. It led to French President and US President endorsing this line of treatment which created favourable public opinion in many countries towards this drug. Liu *et al*. (2020) whose work has also attracted high altmetrics attention found the drug to be effective in “inhibiting SARS-COV-2 in vitro”. This line of treatment and the robustness of the methodology and findings have also generated critical comments, see for example Grens (2020). India has become a key source for this medicine and already exported it to number of countries. Thus a high degree of activity in Google as seen through Google trend can be due to various factors, positive as well as adverse reactions. Similar public opinion generated by studies and endorsement can be seen in YouTube searches in Face Mask.

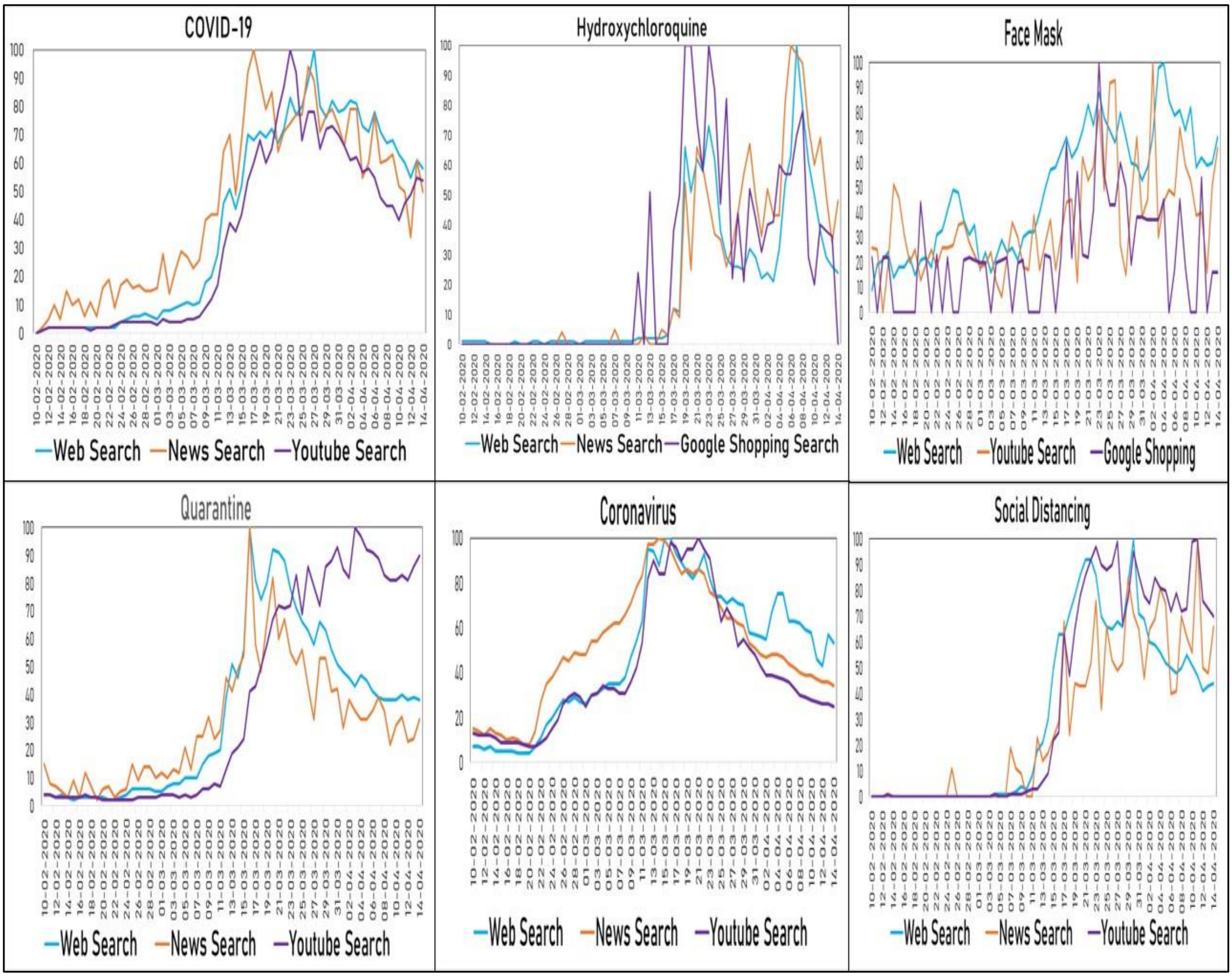

**Figure 3: Globally trending topics related to COVID-19**

Source: Google Trends

Figure 4 provides the comparison of Google trends of “Social Distancing”, “COVID-19”, “Quarantine”, and “Lockdown” and “Face Mask” from 16 Feb to 14 April 2020. Figure 4(a) presents a global picture of these topics and Figure 4 (b) shows data for five countries having maximum cases of COVID-19 namely USA, Italy, Spain, France, Germany and two emerging economies India and Brazil.

**Figure 4: Global Comparison of key trending topics**

**(a) Global Trend**

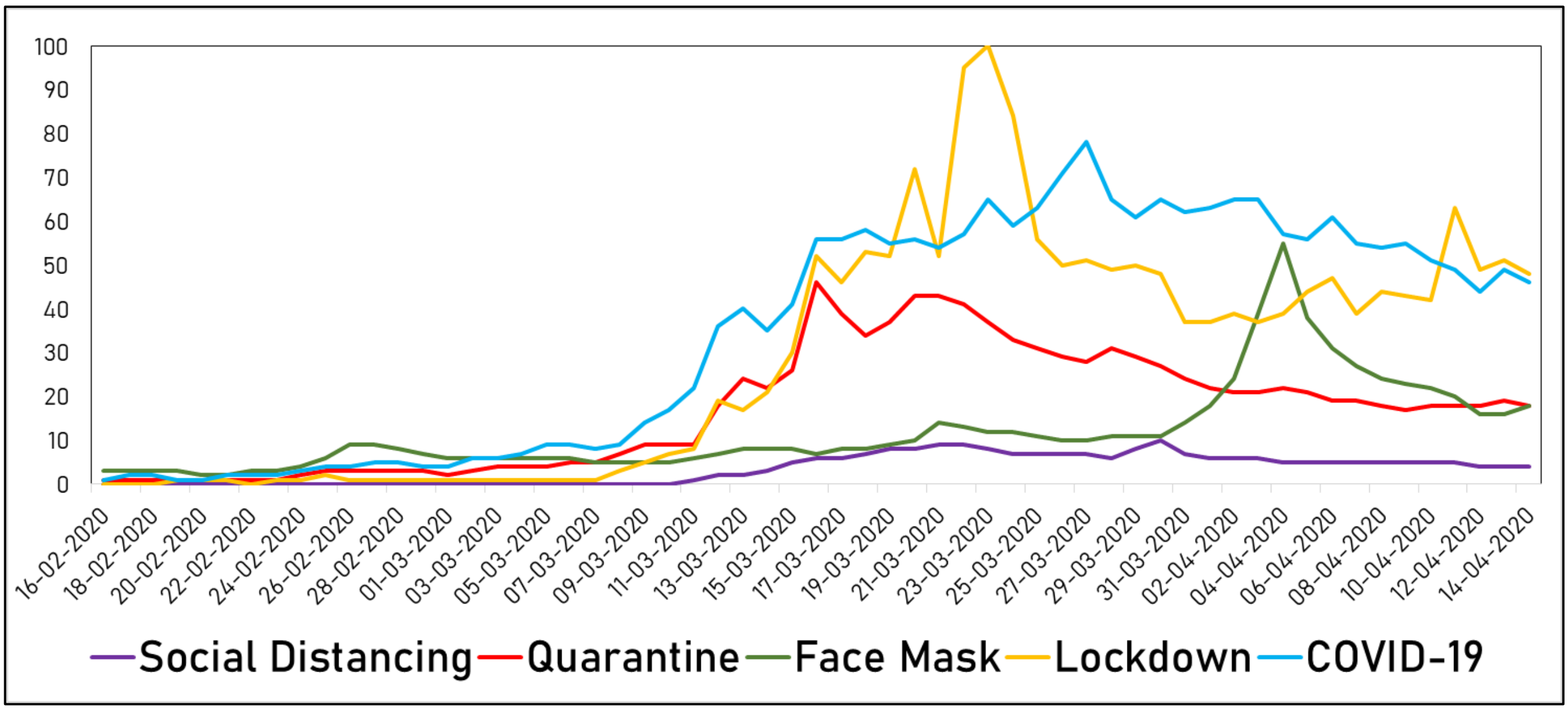

Three of the five topics chosen restrict people's movement. *Social distancing* which basically means keeping a safe distance of around 6 feet from others and avoiding places where this kind of distance cannot be made like schools, workplaces, a sports game or a temple. The second one is *Quarantine* which applies to a person who have been in exposed to coronavirus or patients having coronavirus. The person has to avoid contact with people till the specified incubation period of the virus to see if they develop symptoms. Third *Lockdown*, the term mainly used to describe the confinement of prisoners to their cells has now a changed definition during this outbreak. Through lockdown people are not allowed to leave their local area, building and it is used as a control measure to prevent COVID-19 disease transmission. It can be seen from figure 4 (b) that "Lockdown" is the most popular topic worldwide as well as at the country level. According to World Economic forum 2.6 billion people i.e. one thirds of the world population are under some kind of lockdown.[3] This figure alone constitutes India's 1.3 billion, the largest lockdown in the world. Thus it is unsurprising to see the popularity of this topic over others in India. These control measure have major economic, psychological and social impacts and in turn affects lives of all the people involved. A worrying trend is the low comparative interest in Social Distancing in most of the countries and almost negligible comparative interest in countries like Indian and Brazil. The stand taken by Brazil through her President of opening up and has been critical of measures like social distancing and lockdown may have contributed to this type of trend.

---

[3] https://www.weforum.org/agenda/2020/04/this-is-the-psychological-side-of-the-covid-19-pandemic-that-were-ignoring/

**(b) Country wise Google Web Search Trends**

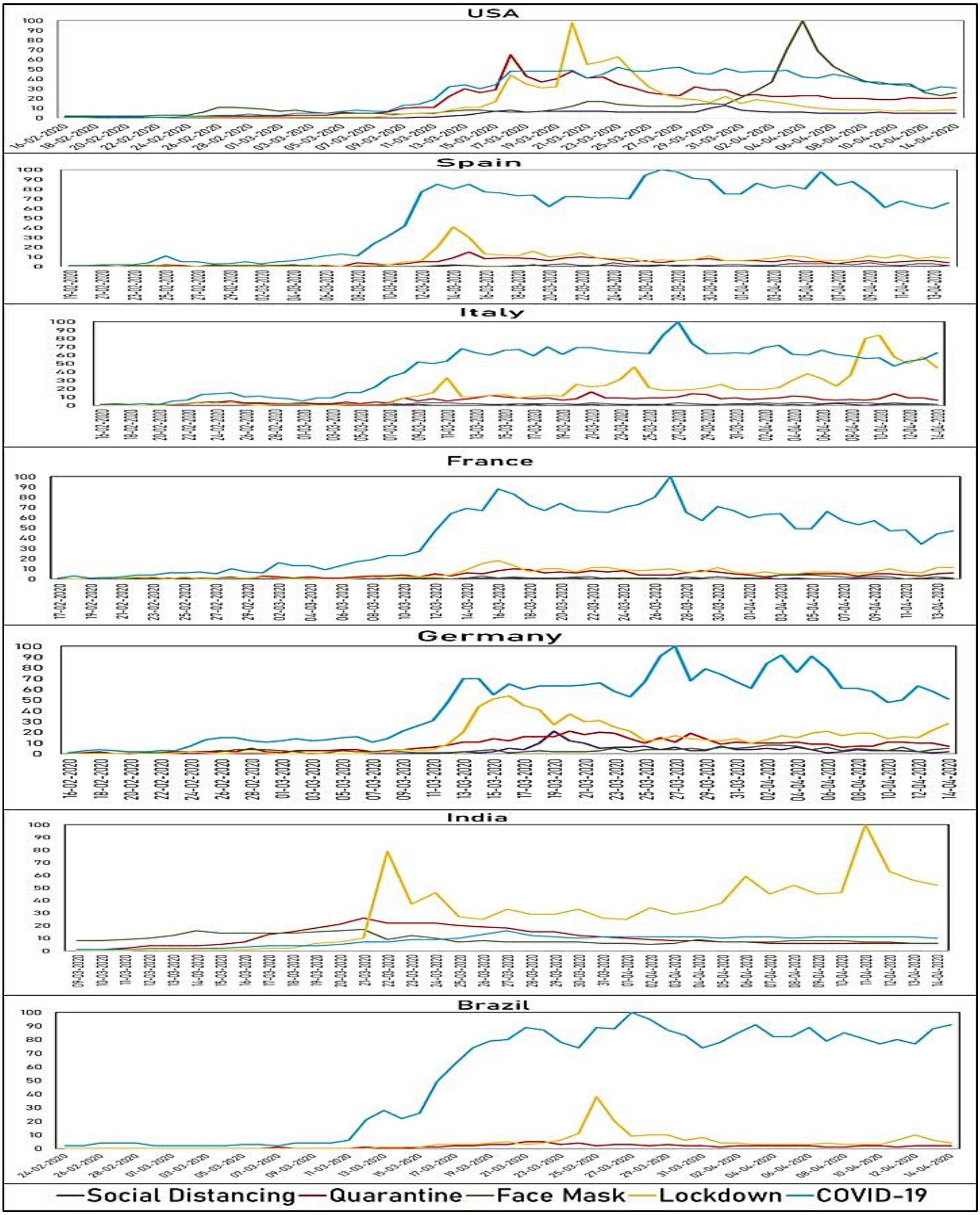

Note: Comparative interest over time of Social Distancing for India and Face Mask and Social Distancing for Brazil were negligible, so these terms have been removed for the Graph for these countries.

## Discussion and Conclusion

COVID-19 disease took the world by surprise. The alarming spread of the disease, challenge to control its spread and its grave health consequences, lack of vaccine or effective drug among others has prompted researchers to actively work on various aspects of this disease. This has led to a huge volume of research output within a short period of time. The various aspects surrounding this disease has also effected society's perception of this disease. Drawing insights from this huge volume of research output is a challenge as well as an exercise of significant importance for the policy makers and research community. Scientometrics provides various tools and techniques to uncover insights from research papers. Altmetrics and Google trends are novel approaches to track online impact. These tools and techniques was used in this study to analyse 9146 papers that were discerned from the Dimensions database covering COVID-19 for the period upto 12th April, 2020. Content analysis of a set of ten key papers was also undertaken to draw qualitative insights of these influential papers.

Some of the insights from this study reveals the key areas in which research has progressed. Their visible impact can be seen in their altmetrics score and more directed impact in their citation in key policy documents. Key policy influence such as period of quarantine, treatment type, use of face masks, population more vulnerable to disease etc can be traced to the influential papers that were discerned from this study. It would be however fallacy of generalisation if we say that these papers were the only influential factors behind the policy decisions. It is also interesting to see how papers had attracted attention from different online sources like twitter, news, blog, and facebook. Thus, the importance of these sources in influencing research impact calls for researchers to aggressively use these modes for dissemination of their study findings.

Another important insight comes from the active collaboration seen in research papers. China and USA drive this research globally and are also actively engaging with other countries in research. This finding is drawn from top ten active countries which constitute almost 78% of the total research output as visible from research papers. Word cloud showed the influential topics of research surrounding COVID-19 research. This type of visual maps provide a good indication of key topics that constitute research activity in a area at a period of time. Co-citation analysis identified the knowledge base that influences current research on this pandemic. It was observed that all of the top co-cited papers were published in high impact factor journals. It was also observed that most of the studies were driven by epidemiology and clinical characteristics of the disease.

Google trends analysis showed how the disease shapes the public opinion on certain topics. Global trending topics incorporated interest over time in web, news, Google shopping and YouTube searches. A sudden rise in a topic's interest could be traced to various exogenous factors. The trend, for example in Hydroxychloroquine, use of this antimalarial drug in the treatment of this disease could be seen the interplay of various forces, the impact of research paper, political endorsement and critical questioning of the research community, doctors etc to the effectiveness of this line of treatment. The trends observed in measures like lockdown, social distancing and quarantine at global and country level showed the societal increasing concern with these aspects.

The findings of this study suggests how the research and public interest has been shaped around this disease. With so much information surrounding this disease, the study provides a space for understanding its various aspects. The study however is limited as it has not examined patents, clinical studies and policy documents. This may provide indications of implementable aspects that draw from research. Future research intends to examine this.

## Acknowledgements

The authors thank Dr Vivek Singh, Professor, Department of Computer Science, Banaras Hindu University for providing access to Dimensions database which helped in framing the pilot study for this research. We are grateful to Dimensions for providing us direct access. Glad to see the initiates they have taken in promoting COVID-19 research.